\documentclass[aps,prd,preprint,nofootinbib]{revtex4}
\usepackage{amsmath,amssymb,graphics,epsfig,subfigure}
\newcommand{\newc}{\newcommand}
\newc{\bsym}{\boldsymbol}
\newc{\mrm}{\mathrm}
\newc{\ovl}{\overline}
\newc{\ovla}{\overleftarrow}
\newc{\ovra}{\overrightarrow}
\newc{\wtil}{\widetilde}
\newc{\eps}{\epsilon}
\newc{\tri}{\triangle}
\newc{\hc}{\dagger}
\newc{\PD}{\partial}
\newc{\ra}{\rightarrow}
\newc{\lra}{\leftrightarrow}
\newc{\rg}{\sqrt{G}}
\newc{\CL}{{\mathcal{L}}}
\newc{\SL}{\!\!\!/}
\newc{\bc}{\begin{center}}
\newc{\ec}{\end{center}}
\newc{\bi}{\begin{itemize}}
\newc{\ei}{\end{itemize}}
\newc{\TR}{\mbox{\sl{Tr}}}
\newc{\hg}{\hat{g}}

\begin{document}
\title{Flavour Changing Neutral Current Constraints from Kaluza-Klein Gluons and Quark Mass Matrices in RS1}
\author{We-Fu Chang}
\email{wfchang@phys.nthu.edu.tw}
\affiliation{Department of Physics, National Tsing Hua University,
Hsin Chu 300, Taiwan}
\author{John N. Ng}
\email{misery@triumf.ca}
\author{Jackson M. S. Wu}
\email{jwu@triumf.ca}
\affiliation{Theory group, TRIUMF, 4004 Wesbrook Mall, Vancouver, B.C., Canada}

\date{\today}

\begin{abstract}
We continue our previous study on what are the allowed forms of quark mass
matrices in the Randall-Sundrum (RS) framework that can reproduce the
experimentally observed quark mass spectrum and the CKM mixing pattern. We
study the constraints the $\Delta F = 2$ processes in the neutral meson sector
placed on the admissible forms found there, and we found only the asymmetrical
type of quark mass matrices arising from anarchical Yukawa structures stay
viable at the few TeV scale reachable at the LHC. We study also the decay of
the first Kaluza-Klein (KK) excitation of the gluon. We give the decay
branching ratios into quark pairs, and we point out that measurements of the
decay width and just one of the quark spins in the dominant $\bar{t}t$ decays
can be used to extract the effective coupling of the first KK gluon to top
quarks for both chiralities. This provides further probe to the flavour
structure of the RS framework.
\end{abstract}


\maketitle

\section{Introduction}
The use of the warped extra-dimensional model of Randall and Sundrum
(RS)~\cite{RS} as a framework for flavour physics has garnered a lot of
attention ever since the model's introduction. By implementing the split
fermion scenario~\cite{SplitF}, the hierarchy in the Standard Model (SM)
fermion masses can be understood geometrically in terms of the different
localization of the SM fermions in the extra dimension~\cite{FermH}. In such a
set-up, the different fermion masses can be obtained without fine tuning the
Yukawa couplings, in contrast to the usual four-dimensional theories.

Having fermions propagating in the extra dimension requires that the SM gauge
symmetry be promoted to a bulk symmetry. Constraints then arise because of the
electroweak precision tests (EWPTs). In particular, for the simplest model with
just the SM gauge group $SU(3)_c \times SU(2)_L \times U(1)_Y$, constraints on
the $S$ and $T$ parameters and the $Zbb$ couplings are found to be difficult
to satisfy without fine tuning. Since an $SU(2)_R$ symmetry is instrumental
in ensuring the very accurate relation $\rho = 1$ in the SM, a natural way to
satisfy the EWPTs would be to promote the $SU(2)_R$ to a bulk gauge symmetry,
and this was done in~\cite{CusRS}.~\footnote{Although having the custodial
symmetry is still the surest way to satisfy the EWPT constraints,
Ref.~\cite{CGHNP08} has reported recently that they may also be satisfied by
having a heavy Higgs boson alone.}

In this work, we continue our study that began in Ref.~\cite{CNW08} of the
forms of quark mass matrices admissible in a minimal RS1 setting with custodial
symmetry that can fit all the experimental data in the quark sector without
having hierarchical Yukawa structures. It is well known that tree-level flavor
changing neutral current (FCNC) interactions are generic in the RS flavour
models. Processes mediated by the Kaluza-Klein (KK) excitations of the gauge
bosons -- in particular that of the gluons -- can give rise to large FCNC
effects, which are tightly constrained by the many low energy measurements in
the neutral meson sector such as $\eps_K$ and $B^0_q$-$\bar{B}^0_q$ transition
($q = d,\,s$). We study in this work the impact these $\Delta F = 2$ FCNC
constraints have on the admissibility of the forms found in Ref.~\cite{CNW08}.
In particular, as these FCNC constraints place stringent limits on the lowest
KK gauge boson mass, $m^{(1)}_{gauge}$, which sets the scale of new physics
(NP), we investigate which of the forms of the quark mass matrices can satisfy
all the FCNC constraints at an NP scale reachable by the LHC. Since the
dominant contribution to the FCNCs comes from the KK gluons in the setting we
study~\footnote{This is explicitly checked in the calculations below.}, we
concentrate on their effects below.

The paper is organized as follow. In Sec.~\ref{sec:MCRS}, we give a brief
outline of the minimal custodial RS (MCRS) model studied to set the notation.
In Sec.~\ref{sec:MesonFV}, we study in the MCRS model the impact of FCNCs
mediated by the tree-level exchange of KK gluons have on the $\Delta F = 2$
processes in the meson sector. These place constraints on the scale of new
flavour physics. In Sec.~\ref{sec:ExptObv}, we evaluate the contribution due
to the KK gluon exchanges in the neutral $B$-meson observables, and we
calculate the branching ratios of the first KK gluon decaying into a pair
of quarks. We point out that measuring even just one of the quark spin, such as
in top decays which are the dominant decay mode, can be very useful in
distinguishing the different models in the RS framework. We conclude in
Sec.~\ref{sec:Concl}. Appendix~\ref{app:ASCfg} contains asymmetrical quark mass
matrices that are typical representations of the families of the admissible
asymmetrical forms used in this work. In Appendix~\ref{app:EWdF}, we show that
with the fermion representation we use in this work, the electroweak
contributions neither displace the dominance of the KK gluon contributions, nor
cause the current FCNC bounds to be violated if they are included as well.

\section{\label{sec:MCRS} The MCRS model}
In this section, we describe briefly the basic set-up of the MCRS model to
establish notations (see also Ref.~\cite{CNW08}) relevant for studying the
flavour changing processes in the meson sector. A more complete and detailed
description can be found in, e.g. Ref.~\cite{CusRS}.

The MCRS mode is formulated on a slice of $AdS_5$ space specified by the metric
\begin{equation}\label{Eq:metric}
ds^2 = G_{AB}\,dx^A dx^B
= e^{-2\sigma(\phi)}\,\eta_{\mu\nu}dx^{\mu}dx^{\nu}-r_c^2 d\phi^2 \,,
\end{equation}
where $\sigma(\phi) = k r_c |\phi|$,
$\eta_{\mu\nu} = \mathrm{diag}(1,-1,-1,-1)$, $k$ is the $AdS_5$ curvature, and
$-\pi\leq\phi\leq\pi$. The theory is compactified on an $S_1/(Z_2 \times Z_2')$
orbifold, with $r_c$ the radius of the compactified fifth dimension, and the
orbifold fixed points at $\phi=0$ and $\phi=\pi$ correspond to the UV (Planck)
and IR (TeV) branes respectively. To solve the hierarchy problem, $k\pi r_c$ is
set to $\approx 37$. The warped down scale is defined to be
$\tilde{k} = k e^{-k\pi r_c}$. Note that $\tilde{k}$ sets the scale of the
first KK gauge boson mass, $m^{(1)}_{gauge} \approx 2.45\tilde{k}$, which
determines the scale of the new KK physics.

The MCRS model has a bulk gauge group
$SU(3)_c \times SU(2)_L \times SU(2)_R \times U(1)_X$ under which the IR
brane-localized Higgs field and transforms as $(1,2,2)_0$. The SM quarks are
embedded into $SU(2)_L \times SU(2)_R \times U(1)_X$ via the five-dimensional
(5D) bulk Dirac spinors
\begin{equation}\label{Eq:qrep}
Q_i =
\begin{pmatrix}
u_{iL}\,[+,+] \\
d_{iL}\,[+,+]
\end{pmatrix} \,,\quad
U_i =
\begin{pmatrix}
u_{iR}\,[+,+] \\
\tilde{d}_{iR}\,[-,+]
\end{pmatrix} \,,\quad
D_i =
\begin{pmatrix}
\tilde{u}_{iR}\,[-,+] \\
d_{iR}\,[+,+]
\end{pmatrix} \,,\qquad
i = 1,\,2,\,3 \,,
\end{equation}
where $Q_i$ transforms as $(2,1)_{1/6}$, and $U_i$, $D_i$ transform as
$(1,2)_{1/6}$. The parity assignment $\pm$ denote the boundary conditions
applied to the spinors on the $[\mathrm {UV}, \mathrm {IR}]$ brane, with $+$
($-$) being the Neumann (Dirichlet) boundary conditions. Only fields with the
[+,+] parity contain zero-modes that do not vanish on the brane. These survive
in the low energy spectrum of the 4D effective theory, and are identified as
the SM fields.

A given 5D bulk fermion field, $\Psi$, can be KK expanded as
\begin{equation}\label{Eq:PsiKK}
\Psi_{L,R}(x,\phi) = \frac{e^{3\sigma/2}}{\sqrt{r_c\pi}}
\sum_{n=0}^\infty\psi^{(n)}_{L,R}(x)f^n_{L,R}(\phi) \,,
\end{equation}
where subscripts $L$ and $R$ label the chirality, and the KK modes $f^n_{L,R}$
are normalized according to
\begin{equation}\label{Eq:fnorm}
\frac{1}{\pi}\int^\pi_{0}\!d\phi\,f^{n\star}_{L,R}(\phi)f^m_{L,R}(\phi)
= \delta_{mn} \,.
\end{equation}
The KK-mode profiles are obtained from solving the equations of motion. For the
zero-modes, the RS flavor functions are  given by
\begin{equation}
f^0_{L,R}(\phi,c_{L,R}) =
\sqrt{\frac{k r_c\pi(1 \mp 2c_{L,R})}{e^{k r_c\pi(1 \mp 2c_{L,R})}-1}}
e^{(1/2 \mp c_{L,R})k r_c\phi} \,,
\end{equation}
where the c-parameter is determined by the bulk Dirac mass parameter,
$m = c\,k$, and the upper (lower) sign applies to the LH (RH) label. Depending
on the orbifold parity of the fermion, one of the chiralities is projected
out.

After spontaneous symmetry breaking, the Yukawa interactions localized on the
IR brane lead to mass terms for the fermions on the IR brane
\begin{equation}
S_\mrm{Yuk} = \int\!d^4x\,\frac{v_W}{k r_c\pi}\Big[
\ovl{\Psi}_u(x,\pi)\lambda^u_{5}\Psi_u(x,\pi)+
\ovl{\Psi}_d(x,\pi)\lambda^d_{5}\Psi_d(x,\pi)\Big]+\mrm{h.\,c.} \,,
\end{equation}
where $v_W = 174$~GeV is the VEV acquired by the Higgs field, and
$\lambda^{u,d}_5$ are the (complex) dimensionless 5D Yukawa coupling matrices.
For zero-modes, this gives the mass matrices for the SM quarks in the 4D
effective theory
\begin{equation}\label{Eq:RSM}
(M^{RS}_f)_{ij} = \frac{v_W}{k r_c\pi}\lambda^f_{5,ij}
f^0_{L}(\pi,c^{L}_{f_i})f^0_{R}(\pi,c^{R}_{f_j})
\equiv \frac{v_W}{k r_c\pi}\lambda^f_{5,ij}F_L(c^{L}_{f_i})F_R(c^{R}_{f_j})
\,, \qquad f = u,\,d \, ,
\end{equation}
where the label $f$ denotes up-type or down-type quark species. The up and
down mass matrices are diagonalized by a bi-unitary transformation
\begin{equation}
(U_L^{u,d})^\hc M^{RS}_{u,d}\,U_R^{u,d} =
\begin{pmatrix}
m^{u,d}_1 & 0         & 0 \\
0         & m^{u,d}_2 & 0 \\
0         & 0         & m^{u,d}_3
\end{pmatrix} \,,
\end{equation}
where $m^{u,d}_i$ are the masses of the SM up-type and down-type quarks. The
mass eigenbasis is then defined by $\psi' = U^\hc\psi$, and the CKM matrix
given by $V_{CKM} = (U^u_L)^\hc U^d_L$.

\section{\label{sec:MesonFV} $\Delta F = 2$ processes in the meson sector}
\begin{figure}[htbp]
\centering
\includegraphics[width=1.5in]{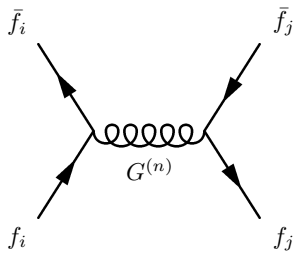}
\caption{\label{Fig:gKK4f} Contributions to $\Delta F = 2$ processes from the
tree-level exchange of KK gluons. The fermions are in the weak eigenbasis.}
\end{figure}

In extra-dimensional models, tree-level flavour changing neutral currents
(FCNCs) arising from the KK-excitations of gauge bosons are generic. For
$\Delta F = 2$ FCNCs, by virtue of the strength of the strong coupling
constant, the largest and thus the most constrained contribution comes from
processes mediated by the exchange of the KK gluons as depicted in
Fig~\ref{Fig:gKK4f}. Effective four-fermion operators are generated when the
KK gluons are integrated out.

In the gauge (weak) eigenbasis, the coupling of the $n$-th level KK gluon,
$G^{(n)}$, to zero-mode fermions is given by
\begin{equation}
G^{(n)}_\mu\left[
\sum_i(g^n_f)^L_{ii}\,\bar{f}_{iL}\gamma^\mu f_{iL} + (L \lra R)\right]
\,, \qquad f = u,\,d \,,
\end{equation}
where $i$ is a generation index, and
$(g^n_f)_{ij} = \mrm{diag}(g^n_{f_1},g^n_{f_2},g^n_{f_3})$ is the weak
eigenbasis coupling matrix with
\begin{equation}
g^n_{f_i} = \frac{g_s}{\pi}\int^\pi_{0}\!d\phi\,
|f^0(\phi,c_{f_i})|^2\chi_n(\phi) \,, \qquad
g_s = \frac{g_{5s}}{\sqrt{r_c\pi}} \,.
\end{equation}
Here, $g_{5s}$ is the bulk 5D $SU(3)$ gauge coupling, $g_s$ that in the SM,
and $\chi_n$ the profile of the n-th KK gluon. Note that the matching relation
between $g_{5s}$ and $g_s$ can be changed by the presence of localized brane
kinetic terms. As in Ref.~\cite{Csaki08}, we have chosen here and for the
analysis below UV boundary terms such that the bare kinetic terms cancel
exactly the contribution coming from the one-loop running. The IR brane kinetic
terms are small and can be neglected.

Going to the mass eigenbasis $f' = U^\hc f$, the $G^{(n)}f'f'$ coupling reads
\begin{equation}
G^{(n)}_\mu\left[
\sum_{a,b}(\hat{g}^n_f)^L_{ab}\,\bar{f}'_{aL}\gamma^\mu f'_{bL}
+(L \lra R)\right] \,, \qquad f = u,\,d \,,
\end{equation}
where
\begin{equation}\label{Eq:gffG}
(\hat{g}^n_f)_{ab}^{L,R} =
\sum_{i,j}(U^\dag_{L,R})_{ai}(g^n_f)_{ij}^{L,R}(U_{L,R})_{jb} \,.
\end{equation}
The off-diagonal couplings $(\hat{g}^n_f)_{ab}$ appear because the diagonal
weak eigenbasis couplings, $g^n_{f_i}$, are not all equal.

In order to compute the coefficients of the effective four-fermion operators
arising from the tree-level KK gluon exchanges, one has to perform (in the
mass eigenbasis) sums of the form
\begin{equation}\label{Eq:KKsum}
\mathfrak{S}^{\omega,\xi}_{ab,cd} = \sum^{\infty}_{n=1}
\frac{(\hat{g}^n_f)^\omega_{ab}(\hat{g}^n_f)^\xi_{cd}}{m_n^2} \,, \qquad
\omega,\,\xi = L,\,R \,,
\end{equation}
where $m_n$ is the mass of the n-th KK gluon, and $\omega$, $\xi$ label the
chirality. The sum over the KK gluon tower can be efficiently calculated with
the help of the massive gauge 5D mixed position-momentum space
propagators~\cite{CusRS,CDPTW03,SR01}~\footnote{See also Ref.~\cite{Csaki08}
for an equivalent way of summing up the gluon KK tower.}. It can be computed
in terms of the overlap integral,
\begin{equation}
G^{++}_{ff}(c^\omega_i,c^\xi_j) = \frac{1}{\pi}\!\int_0^{\pi}\!d\phi
|f^0_\omega(\phi,c^\omega_i)|^2\tilde{G}^{(++)}_{p=0}(\phi,\phi')
|f^0_\xi(\phi',c^\xi_j)|^2 \,, \qquad \omega,\,\xi = L,\,R \,,
\end{equation}
where $\tilde{G}^{(++)}_{p=0}$ is the zero-mode subtracted gauge propagator
evaluated at zero 4D momentum, and is given by~\cite{CDPTW03}
\begin{align}
\tilde{G}^{(++)}_{p=0}(\phi,\phi') = \frac{1}{4k(k r_c\pi)}\bigg\{
\frac{1-e^{2k r_c\pi}}{k r_c\pi}+e^{2k r_c\phi_<}(1-2k r_c\phi_<)
+e^{2k r_c\phi_>}\Big[1+2k r_c(\pi-\phi_>)\Big]\bigg\} \,,
\end{align}
where $\phi_< = \mrm{min}(\phi,\phi')$, $\phi_> = \mrm{max}(\phi,\phi')$. The
sum over the KK tower is then given by
\begin{equation}\label{Eq:g4f}
\mathfrak{S}^{\omega,\xi}_{ab,cd} = g_s^2\sum_{i,j}
(U^\hc_\omega)_{ai}(U_\xi)_{ib}\,G^{++}_{ff}(c^\omega_i,c^\xi_j)\,
(U^\hc_\omega)_{cj}(U_\xi)_{jd} \,, \qquad \omega,\,\xi = L,\,R \,.
\end{equation}

The most general effective Hamiltonian for the $\Delta F = 2$ processes beyond
the SM can be written as
\begin{equation}
\mathcal{H}^\mrm{NP}_\mrm{eff} =
\sum_{i=1}^5 C_i(\Lambda)Q_i^{ab} +
\sum_{i=1}^3\wtil{C}_i(\Lambda)\wtil{Q}_i^{ab} \,,
\end{equation}
where $\Lambda$ is the scale of new physics (NP), and
\begin{align}
Q_1^{ab} &=
\bar{\psi}^\alpha_{aL}\gamma_\mu\psi^\alpha_{bL}
\bar{\psi}^\beta_{aL}\gamma^\mu\psi^\beta_{bL} \,, \notag \\
Q_2^{ab} &=
\bar{\psi}^\alpha_{aR}\psi^\alpha_{bL}
\bar{\psi}^\beta_{aR}\psi^\beta_{bL} \,, \notag \\
Q_3^{ab} &=
\bar{\psi}^\alpha_{aR}\psi^\beta_{bL}
\bar{\psi}^\beta_{aR}\psi^\alpha_{bL} \,, \notag \\
Q_4^{ab} &=
\bar{\psi}^\alpha_{aR}\psi^\alpha_{bL}
\bar{\psi}^\beta_{aL}\psi^\beta_{bR} \,, \notag \\
Q_5^{ab} &=
\bar{\psi}^\alpha_{aR}\psi^\beta_{bL}
\bar{\psi}^\beta_{aL}\psi^\alpha_{bR} \,,
\end{align}
with $\alpha$, $\beta$ the colour indices, and $a,b$ the generation
indices~\footnote{The so-called supersymmetric (SUSY) basis of
operators~\cite{GGMS96} is used here. Other basis can be obtained via the
appropriate Fierz identities.}. The operators $\wtil{Q}^{ab}_{1,2,3}$ are
obtained from $Q^{ab}_{1,2,3}$ by the $L \lra R$ exchange. All operators are
given in the mass eigenbasis here. In the MCRS model, only $Q^{ab}_{1,4,5}$
and $\wtil{Q}^{ab}_1$ arise from the tree-level exchange of KK gluons, and
their coefficients are given by
\begin{equation}\label{Eq:WCs}
C_1(\Lambda) = \frac{1}{6}\mathfrak{S}^{LL}_{ab,ab} \,, \qquad
\wtil{C}_1(\Lambda) = \frac{1}{6}\mathfrak{S}^{RR}_{ab,ab} \,, \qquad
C_4(\Lambda) = -\mathfrak{S}^{LR}_{ab,ab} \,, \qquad
C_5(\Lambda) = -\frac{1}{3}C_4 \,.
\end{equation}
Note that here the NP scale is the scale where the KK excitations first come
in, hence $\Lambda \sim m_1$.

Recently, a model independent global analysis of the physical observables in
the $\Delta F = 2$ processes have been performed by the UTfit
collaboration~\cite{UTfit07}. Bounds on the NP scale Wilson coefficients
$C_i(\Lambda)$ above have been found with the Renormalization Group evolution
fully taken into account. Given these bounds, an immediate question with regard
to the admissible forms of quark mass matrices found in Ref.~\cite{CNW08} is
whether they remain viable, as they govern the form of the rotation matrices,
$U_\omega$, that determine the Wilson coefficients in the MCRS model (see
Eqs.~\eqref{Eq:g4f} and~\eqref{Eq:WCs}).

Two types of mass matrix structures were found in Ref.~\cite{CNW08} that
reproduce well the observed patterns of quark masses and CKM mixings, and are
compatible with non-hierarchical and perturbative Yukawa structures
($|\lambda_5| < 4$~\cite{APS05}) in the RS framework. In one type, mass
matrices have a symmetrical texture that is a slight deformation of the ansatz
proposed by Koide~\textit{et. al.}~\cite{KNMKF02}. In the other, there are no
symmetries \textit{a priori}. The form of the mass matrices is characterized by
the localizations of the fermions in the 5D bulk that are admissible under the
electroweak constraints, and each particular realization of the form arise from
Yukawa structures that are completely anarchical. For each type of the quark
mass matrices, we calculate below the resulting Wilson coefficients for the
$\Delta F = 2$ processes due to KK gluon exchanges, and we compare them to the
UTfit bounds.

For the symmetrical Koide-type form of quark mass matrix, we begin by focusing
on the kaon sector where the constraints are most stringent~\cite{UTfit07}. At
$\Lambda = 4$~TeV, while the imaginary part of the resulting kaon sector Wilson
coefficients are all very much smaller than the bounds listed, we find the
real parts are all larger than the respective bounds by three orders of
magnitude. As a result, insisting that the symmetrical type pass the UTfit
bounds would require one to push the NP scale up to
$\mathcal{O}(100)$~TeV~\footnote{As can be seen from  Eq.~\eqref{Eq:KKsum}, the
mass of the lightest mode sets the suppression scale for the four-fermion
operator. To make up for a factor of $\mathcal{O}(10^3)$ at
$m_1\simeq\Lambda = 4$~TeV would require a factor of $\mathcal{O}(30)$
increase in $m_1$.}.

For asymmetrical forms, we demonstrate that each of the asymmetrical
configurations discussed in Ref.~\cite{CNW08} remain viable at the few TeV
scale. In Table~\ref{Tb:WCFC}, we list the UTfit bounds on the relevant Wilson
coefficients, and we give their values for a typical ``solution'' -- admissible
set of up and down quark mass matrices which give the observed quark masses
and mixings, and satisfy all electroweak and FCNC bounds -- at
$\Lambda = 4$~TeV (corresponding to $\tilde{k} = 1.65$~TeV where
$m_1 \simeq 4$~TeV) in each of the asymmetrical configurations, and we see that
they are all well within the bounds. The details of the specific quark mass
matrices used are given in Appendix~\ref{app:ASCfg}. In all calculations, we
have explicitly checked that the KK gluons do indeed give rise to the dominant
contributions in the tree-level $\Delta F = 2$ process under study. We show in
Appendix~\ref{app:EWdF} that the contributions from the electroweak sector are
small as expected, and would not lead to violations of the UTfit bounds if
included with the KK gluon contributions.

\begin{table}[htbp]
\begin{ruledtabular}
\begin{tabular}{ccccc}
Parameter & 95\% allowed range & Config. I & Config. II & Config. III \\
\hline
Re $C^1_K$ & $[-9.6,9.6] \cdot 10^{-13}$ &
$4.3 \cdot 10^{-17}$ & $1.8 \cdot 10^{-15}$  & $-4.2 \cdot 10^{-15}$ \\
Re $C^4_K$ & $[-3.6,3.6] \cdot 10^{-15}$ &
$-1.4 \cdot 10^{-16}$ & $-2.8 \cdot 10^{-16}$ & $-1.8 \cdot 10^{-15}$ \\
Re $C^5_K$ & $[-1.0,1.0] \cdot 10^{-14}$ &
$4.6 \cdot 10^{-17}$& $9.4 \cdot 10^{-17}$  & $6.0 \cdot 10^{-16}$ \\
\hline
Im $C^1_K$ & $[-4.4,2.8] \cdot 10^{-15}$ &
$2.6 \cdot 10^{-18}$ & $1.8 \cdot 10^{-15}$  & $-1.0 \cdot 10^{-15}$ \\
Im $C^4_K$ & $[-1.8,0.9] \cdot 10^{-17}$ &
$1.5 \cdot 10^{-19}$ & $8.8 \cdot 10^{-18}$  & $-1.8 \cdot 10^{-18}$ \\
Im $C^5_K$ & $[-5.2,2.8] \cdot 10^{-17}$ &
$-4.9 \cdot 10^{-20}$ & $-2.9 \cdot 10^{-18}$ & $6.0 \cdot 10^{-19}$ \\
\hline
$|C^1_D|$ & $< 7.2 \cdot 10^{-13}$ &
$1.3 \cdot 10^{-13}$ & $3.1 \cdot 10^{-13}$ & $1.6 \cdot 10^{-14}$ \\
$|C^4_D|$ & $< 4.8 \cdot 10^{-14}$ &
$1.7 \cdot 10^{-15}$ & $8.8 \cdot 10^{-15}$ & $4.0 \cdot 10^{-14}$ \\
$|C^5_D|$ & $< 4.8 \cdot 10^{-13}$ &
$5.7 \cdot 10^{-16}$ & $2.9 \cdot 10^{-15}$ & $1.3 \cdot 10^{-14}$ \\
\hline
$|C^1_{B_d}|$ & $< 2.3 \cdot 10^{-11}$ &
$7.5 \cdot 10^{-13}$ & $7.7 \cdot 10^{-14}$ & $4.8 \cdot 10^{-13}$ \\
$|C^4_{B_d}|$ & $< 2.1 \cdot 10^{-13}$ &
$1.9 \cdot 10^{-13}$ & $4.8 \cdot 10^{-14}$ & $1.7 \cdot 10^{-13}$ \\
$|C^5_{B_d}|$ & $< 6.0 \cdot 10^{-13}$ &
$6.2 \cdot 10^{-14}$ & $1.6 \cdot 10^{-14}$ & $5.6 \cdot 10^{-14}$ \\
\hline
$|C^1_{B_s}|$ & $< 1.1 \cdot 10^{-9}$  &
$9.0 \cdot 10^{-11}$ & $4.1 \cdot 10^{-11}$ & $4.0 \cdot 10^{-11}$ \\
$|C^4_{B_s}|$ & $< 1.6 \cdot 10^{-11}$ &
$9.4 \cdot 10^{-12}$ & $7.6 \cdot 10^{-13}$ & $5.8 \cdot 10^{-12}$ \\
$|C^5_{B_s}|$ & $< 4.5 \cdot 10^{-11}$ &
$3.1 \cdot 10^{-12}$ & $2.5 \cdot 10^{-13}$ & $1.9 \cdot 10^{-12}$ \\
\end{tabular}
\end{ruledtabular}
\caption{\label{Tb:WCFC} The 95\% allowed range of the Wilson
coefficients~\cite{UTfit07} contributing in the $\Delta F = 2$ tree-level
gluon exchange processes, and their typical values at $\Lambda = 4$~TeV in each
of the asymmetrical configurations given in Ref.~\cite{CNW08}. All values above
are given in units of $\mrm{GeV}^{-2}$.}
\end{table}

Note that in Table~\ref{Tb:WCFC}, only one of the many admissible solutions we
found are given for each asymmetrical configurations. Moreover, the
configurations of fermion localizations themselves are just three of many that
we found which lead to admissible solutions. Indeed, we have found that
parameter space generically exists in the RS1 setting where quark mass and
mixing data and $\Delta F = 2$ FCNC bounds can be satisfied at the few TeV
scale with asymmetrical quark mass matrices that arise from underlying
anarchical Yukawa structures. This does not, however, contravene the conclusion
reached in Ref.~\cite{Csaki08} that a KK scale of 10 to 20 TeV is necessary to
satisfy the $\Delta F = 2$ FCNC constraints in the kaon sector. The higher NP
scale is required if one wants to ensure that the FCNC bounds is generically
satisfied for any given quark mass matrices that give the pattern of the
observed quark mass hierarchy and CKM mixings. Our point here is that a subset
of these consisting of asymmetrical quark mass matrices exists such that the
experimental quark masses and mixings are reproduced to a high accuracy, and
at the same time the lower, few TeV scale is still viable.~\footnote{The few
TeV NP scale can also be achieved one imposes additional symmetries. For recent
works in this direction see e.g. Refs.~\cite{S08,CFW08}.} We emphasize here that 
this subset does not contain isolated singular points in the parameter space, 
but generic solutions throughout all the parameters space.

Now one may worry that radiative correction may spoil our results, as there are
loop induced corrections to the brane localized Yukawa couplings, and loop
induced brane kinetic mixing terms that can introduce additional flavour
violations. This is however not so. First, we are not calculating theoretically
the Yukawa couplings in the RS framework; to do that requires a UV completion
of the theory. There will be radiative corrections to the Yukawa matrices, but
they will not change the form of the 4D effective mass matrices given in
Eq.~\eqref{Eq:RSM} even if it is derived at tree-level. Thus if the physical
(or renormalized) Yukawa matrices take any of the forms that we found, the
FCNC bounds will be satisfied, the form of the mass matrices we give should
therefore be viewed as ``physical'' and the corresponding Yukawa matrices
renormalised. Next, the brane kinetic mixings, which is loop suppressed, lead
to a correction to the gauge-fermion couplings of order
$\delta\sim|\lambda_{5D}|^2/4\pi^2$ as can be estimated from NDA (naive
dimensional analysis)~\cite{CFW08}. In the search for solutions, we have set
$|\lambda_{5D}| \lesssim 2$, consequently $\delta \ll 1$ and the flavour
violating contribution from the brane kinetic mixing terms is small
($\mathcal{O}(0.01)$ of the KK gluon contributions), which do not impact the
Wilson coefficients calculated.

\section{\label{sec:ExptObv} Experimental observables}
\subsection{$B^0_q$-$\bar{B}^0_q$ mixings}
One very sensitive probe to NP in the meson sector comes from the
$B^0_q$-$\bar{B}^0_q$ mixing ($q = d,\,s$), which has received much theoretical
attention, and has now an extensive body of experimental data from the $B$
factories and FNAL. The contribution of NP to $\Delta B = 2$ transitions can
be parametrized in a model-independent way as the ratio of the full (SM + NP)
amplitude to the SM one~\cite{UTfit07}~\footnote{See also Ref.~\cite{BF06}.}:
\begin{equation}
\frac{\langle B^0_q|\mathcal{H}^\mrm{full}_\mrm{eff}|\bar{B}^0_q\rangle}
{\langle B^0_q|\mathcal{H}^\mrm{SM}_\mrm{eff}|\bar{B}^0_q\rangle} = 1 +
\frac{\langle B^0_q|\mathcal{H}^\mrm{NP}_\mrm{eff}|\bar{B}^0_q\rangle}
{\langle B^0_q|\mathcal{H}^\mrm{SM}_\mrm{eff}|\bar{B}^0_q\rangle}\equiv
C_q\,e^{2i\phi_q} \,, \qquad q = d,\,s \,,
\end{equation}

The SM amplitude arise mainly from the one-loop box diagram, which is
dominated by the top quark exchanges. It can be written as
\begin{equation}
\langle B^0_q|\mathcal{H}^\mrm{SM}_\mrm{eff}|\bar{B}^0_q\rangle =
\frac{G_F^2 m_W^2}{6\pi^2}\hat{\eta}_B m_{B_q}^2 f_{B_q}^2\hat{B}_{B_q}
(V_{tq}^\ast V_{tb})^2 S_0(x_t) \,, \qquad
x_t\equiv\frac{m_t^2}{m_W^2} \,,
\end{equation}
where $\eta_B = 0.552$ is a short distance QCD correction~\cite{BBL95}, and
$S_0$ is an ``Inami-Lim'' function~\cite{IL80} with
$m_t(m_t) = 163.6$~GeV~\cite{XZZ08}. We take for the CKM mixings~\cite{BF06}
\begin{equation}
|V_{td}^\ast V_{tb}| = 8.6 \cdot 10^{-3} \,, \qquad
|V_{ts}^\ast V_{tb}| = 41.3 \cdot 10^{-3} \,,
\end{equation}
for the decay constants~\cite{fB07}
\begin{equation}
f_{B_d} = 197\,\mrm{MeV} \,, \qquad f_{B_s} = 240\,\mrm{MeV} \,.
\end{equation}
and for the renormalization invariant bag parameter~\cite{Okamoto06}
\begin{equation}
f_{B_d}\sqrt{\hat{B}_{B_d}} = 244\,\mrm{MeV} \,, \qquad
f_{B_s}\sqrt{\hat{B}_{B_d}} = 295\,\mrm{MeV} \,,
\end{equation}
All other input parameters take their values from the PDG~\cite{PDG08}.

In the MCRS model, the NP contribution to the $\Delta B = 2$ transition
amplitude is dominated by the tree-level exchanges of KK gluons, as the
coupling strength for the strong interactions is much larger than that for the
electroweak interactions. Evolving down from the NP scale $\Lambda$ to the
hadronic scale $\mu = m_b$, the KK gluon contribution is given by
\begin{equation}\label{Eq:NPampl}
\langle B^0_q|\mathcal{H}^\mrm{NP}_\mrm{eff}|\bar{B}^0_q\rangle =
\langle B^0_q|
\sum_r C_r(\mu)Q^{bq}_r(\mu) + \sum_s\wtil{C}_s(\mu)\wtil{Q}^{bq}_s(\mu)
|\bar{B}^0_q\rangle \,,
\end{equation}
where
\begin{equation}
C_r(\mu) = \sum_{i,j}\left(
b^{(r,i)}_j+\eta\,c^{(r,i)}_j\right)\eta^{\alpha_j}C_i(\Lambda) \,, \qquad
\eta = \frac{\alpha_s(\Lambda)}{\alpha_s(m_t^{pole})} \,,
\end{equation}
are the Wilson coefficients at the hadronic scale, with $\wtil{C}_r$ defined
similarly with the same coefficients as for $C_r$, and
$m_t^{pole} = 171.4$~GeV~\cite{PDG08}. The magic numbers $\alpha_j$,
$b^{(r,i)}_j$, $c^{(r,i)}_j$, and the operator matrix elements can be found in
Ref.~\cite{Bmagic01}~\footnote{Note that Ref.~\cite{Bmagic01} works in the
Landau RI-MOM scheme~\cite{RIMOM}; for magic numbers in the $\ovl{\mrm{MS}}$
(NDR) scheme, see Ref.~\cite{BJR01}. For consistency, all running quark masses
used in Eq.~\eqref{Eq:NPampl} should be in the same scheme as the operator
matrix elements. The relevant quark masses in the RI-MOM scheme are
$m_b(m_b) = 4.6$~GeV, $m_s(m_b) = 87$~MeV, and $m_d(m_b) = 5.4$~MeV.}.

In Table~\ref{Tb:B0NP}, we give the values of the parameters $C_q$ and
$\phi_q$ for each of the three asymmetrical configurations solutions used in
Table.~\ref{Tb:WCFC}.
\begin{table}[htbp]
\begin{ruledtabular}
\begin{tabular}{cccc}
Parameter & Config. I & Config. II & Config. III \\
\hline
$C_d$ & 1.13 & 1.02 & 1.08 \\
$\phi_d\,[^\circ]$ & -2.48 & -0.24 & -3.02 \\
$C_s$ & 1.68 & 1.36 & 1.29 \\
$\phi_s\,[^\circ]$ & 0.61 & 0.12 & 0.04 \\
\end{tabular}
\end{ruledtabular}
\caption{\label{Tb:B0NP} Parameters determining the NP contributions to
$B^0_q$-$\bar{B}^0_q$ mixings in the MCRS model with mass matrices from the
three asymmetrical configurations given in Ref.~\cite{CNW08}.}
\end{table}
The values of $C_q$ and $\phi_q$ agree well with the UTfit values at 95\%
probability (and mostly at 68\% as well; see Table~3 in Ref.~\cite{UTfit07})
as expected, since the physical observables fitted here are the same that go
into the analysis for the meson sector flavour bound on the NP Wilson
coefficients listed in Table~\ref{Tb:WCFC}. As above, we have checked that the
electroweak contributions is small -- they are much less than the standard
error given by the UTfit collaboration at 68\% probability -- and do not cause
large shifts that would violate the UTfit bounds. We note for the
configurations of solutions given here, KK gluons are not manifest in the
$B^0_q$-$\bar{B}^0_q$ mixing, and the SM effects are expected to be dominant.

\subsection{KK gluon top decays}
In RS models, a distinguishing property of the KK gluons is that their
couplings to the LH and RH fermions (in the mass eigenbasis), $\hat{g}_L$ and
$\hat{g}_R$, are in general not the same. For all the asymmetrical quark mass
matrix solutions that we found, this is true. To test this experimentally, one
way is to measuring both the decay width and the spin of the top in the decays
of gluons into top pairs as we show below. We will also be concentrating on
the first KK gluon, $G^{(1)}$, which has the highest potential of being within
the reach of the LHC.

As $G^{(1)}$ couples strongly to states localized near the IR brane, and the
large top mass requires that either $Q_3$ or $t_R$ be IR localized, top decays
are expected to be dominant modes of decay. The partial width of $G^{(1)}$
decaying into quarks in the mass eigenbasis, $\bar{q}_a q_b$, can be written as
\begin{align}
\Gamma(G^{(1)}\ra\bar{q}_a q_b) &=
\frac{m_1}{48\pi}\lambda(1,x_a^2,x_b^2)^{1/2} \notag \\
&\qquad\times\Big\{\frac{1}{2}(|\hg_L^1|^2 + |\hg_R^1|^2)
\left[2-x_a^2-x_b^2-(x_a^2-x_b^2)^2\right]
+6\,\mrm{Re}[\hg_L^1(\hg_R^1)^\ast]\,x_a x_b\Big\} \notag \\
&\ra\frac{m_1}{48\pi}(|\hg_L^1|^2 +|\hg_R^1|^2) \quad (x_a = x_b \ll 1) \,,
\qquad x_{a,b} = \frac{m_{a,b}}{m_1} \,,
\end{align}
where $m_1$ is the mass of $G^{(1)}$, $\hat{g}^1_{L,R}$ denote the mass
eigenbasis couplings, $(\hat{g}^1_f)^{L,R}_{ab}$, given in Eq.~\eqref{Eq:gffG},
and $\lambda(u,v,w) = (u-v-w)^2-4vw$. For the three asymmetrical configuration
solutions used in Table~\ref{Tb:WCFC}, and for $m_1 = 4.0$~TeV, the widths into
the $\bar{t}t$ pairs are:
\begin{equation}\label{Eq:twidth}
769.3~\mrm{GeV} \; (\mrm{Config.\,I}) \,, \qquad
635.4~\mrm{GeV} \; (\mrm{Config,\,II}) \,, \qquad
747.4~\mrm{GeV} \; (\mrm{Config.\,III}) \,.
\end{equation}
In Table~\ref{Tb:G1Br}, we give the branching ratios of $G^{(1)}$ into top,
bottom, and all other modes involving at least on light quark (jets) for the
same three asymmetrical solutions.
\begin{table}[htbp]
\begin{ruledtabular}
\begin{tabular}{cccc}
Branching ratios & Config. I & Config. II & Config. III \\
\hline
Top quarks & 0.83 & 0.83 & 0.84 \\
Bottom quarks & 0.16 & 0.16 & 0.15 \\
Light quarks & 0.01 & 0.01 & 0.01 \\
\end{tabular}
\end{ruledtabular}
\caption{\label{Tb:G1Br} Branching ratios of $G^{(1)}$ into $\bar{q}_a q_b$
pair in the MCRS model with mass matrices from the three asymmetrical
configurations given in Ref.~\cite{CNW08}. The term ``light quarks'' here
denotes all modes (flavour changing included) that involve at least one light
quark (jet).}
\end{table}

We see from Table~\ref{Tb:G1Br} that most decays are into top pairs, with
negligible fraction into light quarks. Compared to Ref.~\cite{KKLHC} (see
Table~I), the branching ratio into top pairs from each of our asymmetrical
configurations is slightly lower at about 80\% instead of around 90\%, which
is due to the difference in the quark mass matrices and the localization
parameters used. Note that the branching ratios are stable across the
different configurations. This is because the couplings of KK gluons to quarks
are dominated by that to the third generation quarks, which varied little
across the configurations. It is thus fairly robust that in the RS scenario,
the KK gluon will decay predominantly into top pairs, and then into b-jets
with a much smaller, but non-negligible rate. Other light quark modes are
negligible and certainly no leptons. However, as can be seen from
Eq.~\eqref{Eq:twidth}, the top pair width is not small as
$\Gamma_{\bar{t}t}/m_1 \sim 0.2$. Thus looking for signals in the resonant
productions will require good top identification. Detail discussions of the
discovery potential at the LHC can be found in Ref.~\cite{KKLHC}. We note that
the bottom mode should not be overlook and can be used as a check if not the
primary discovery tool.

If a KK gluon is found at the LHC, it will certainly be important to measure
the spin of the top in its $\bar{t}t$ decays. The differential decay rate with
only one of the top spin measured but with the other top spin summed over is
given by
\begin{align}
\frac{d\Gamma_s}{d\cos{\theta}} = \frac{m_1}{192\pi}\sqrt{1-4x_t^2}\Big\{
&(|\hg_L|^2 + |\hg_R|^2)(1-x_t^2) + 6\,\mrm{Re}\,(\hg_L\hg_R^\ast)\,x_t^2
\notag \\
&+ 2(|\hg_R|^2-|\hg_L|^2)\,x_t\sqrt{1-4x_t^2}\,
\mathbf{s}\cdot\mathbf{\hat{p}}\Big\} \,,
\end{align}
where
\begin{equation}
\mathbf{s}\cdot\mathbf{\hat{p}} =
\frac{\cos{\theta}}{\sqrt{1-(1-4x_t^2)\cos^2\theta}} \,, \qquad
\cos\theta\equiv\mathbf{\hat{s}}\cdot\mathbf{\hat{p}} \,,
\end{equation}
with $\mathbf{s}$ the measured top spin three-vector, and $\mathbf{p}$ the
three-momentum of the same top quark in the rest frame of $G^{(1)}$. From this
we see that a measurement of the angular dependence together with the decay
width can allow one to extract out $\hg_L$ and $\hg_R$. The feasibility of
doing this at the LHC requires detailed numerical simulations which are beyond
the scope of the present work (see Ref.~\cite{Atlast} for work in this
direction).

\section{\label{sec:Concl} Conclusion}
Previously in Ref.~\cite{CNW08} we have studied the phenomenologically allowed
form of quark mass matrices in the MCRS model, and we have found admissible
both a symmetrical form, and many distinctive asymmetrical configurations with
Yukawa structures non-hierarchical and anarchical that satisfy all EWPTs. The
benchmark warped down scale was chosen at $\tilde{k} = 1.65$~TeV implying an
equivalently NP scale of $\Lambda = 4$~TeV, since a higher scale will prevent
the KK gauge bosons being detectable at the LHC initially at least. A much
higher scale will also create its own hierarchy problem which one would like
to avoid. We continue the study in this work the constraints that
$\Delta F = 2$ processes in the neutral meson sector impose. We found from
these constraints that for the symmetrical mass quark matrices, the viable
scale is pushed up to $\mathcal{O}(100)$~TeV. However, for the asymmetrical
quark mass matrices, $\Lambda = 4$~TeV is still viable. This is a consequence
of the fact that in the asymmetrical cases there is freedom in the LH and RH
rotations being very different -- rather than being locked into a specific
pattern as in the symmetrical case -- which can supply the suppression
required to pass the meson sector $\Delta F = 2$ constraints. This underscores
the importance of the quark mass matrices in the RS framework both
phenomenologically and theoretically for identifying any family symmetries that
may be hidden.

At the $\Lambda = 4$~TeV scale, discovery of the first KK gluon state at the
LHC is possible. This can be achieved through a resonance search in the
$\bar{t}t$ channel which we predict to have a branching ratio of $\approx 0.8$.
Note that the dominance of the $\bar{t}t$ decays is a characteristic of the
RS1 scenario. The $\bar{b}b$ mode has a branching fraction of about 0.15, and
should not be overlooked. This mode consists mainly of LH pairs because $b_L$
is an $SU(2)$ partner to $t_L$, which has a large overlap with $G_{KK}$. Thus
this channel can be useful as a diagnostic tool if the expected background can
be handled. All other decay modes involving light quarks are negligible.
Finally, if one can also measure in the $\bar{t}t$ decays at least one of the
quark spins, it will help to unravel $\hg_L$ and $\hg_R$, and provide further
an invaluable probing into the flavour structure of the RS scenario.

\section{acknowledgments}
We thank C. Cs\'{a}ki for useful comments.
W.F.C. is grateful to the TRIUMF Theory group for their
hospitality when part of this work was completed. The research of
J.N.N. and J.M.S.W. is partially supported by the Natural Science
and Engineering Council of Canada. The work of W.F.C. is supported
by the Taiwan NSC under Grant No. 96-2112-M-007-020-MY3.

{\em Note added}: At the time when this work was completed, Ref.~\cite{BBDGW08}
came out which also considered some of the same issues as us. There, the bulk
gauge group contains an additional discrete left-right parity group. Consequently
the fermion matter contents are embedded in a different representation than the
one in our work, resulting in electroweak contributions to the $\Delta F = 2$
FCNCs that are far larger than what we have found. We have checked that both works
agree whenever direct comparisons can be made.

\appendix

\section{\label{app:ASCfg} Typical solutions for the asymmetrical
configurations}
In this appendix, we give the details of the quark mass matrices of the
typical solution used in Table~\ref{Tb:WCFC} in each of the three asymmetrical
configurations given in Ref.~\cite{CNW08}. Although the bound on $Z\bar{b}_L b_L$
used there was that given in the PDG~\cite{PDG08}, many generic solutions from 
generic configurations exist with localization parameters that can easily 
accommodate the more stringent bound found in e.g. Ref.~\cite{AC06}.

Parameterizing the complex 5D Yukawa couplings as
$\lambda_{5,ij} = \rho_{ij}e^{i\phi_{ij}}$, admissible mass matrices of the
forms given by Eq.~\eqref{Eq:RSM} are found with $\rho_{ij}$ and $\phi_{ij}$
randomly and uniformly generated in the intervals $(0,2)$ and $[-\pi,\pi)$
respectively. In the following, we list the complex mass matrices in the form
of $M_f = |M_f|e^{i\theta_f}$, the magnitude and phase of the 5D Yukawa
couplings, and the mass eigenvalues for both the up and down sector. All values
are given to six significant figures. The mass eigenvalues agree with the
quark masses at 1~TeV found in Ref.~\cite{XZZ08} to within two standard
deviations quoted.

\begin{itemize}
\item Configuration~I:
\begin{align}
c_Q &= \{0.633604, 0.556171, 0.256293\} \,, \notag \\
c_U &= \{-0.663816, -0.535621, 0.185413\} \,, \notag \\
c_D &= \{-0.641469, -0.572479, -0.616085\} \,.
\end{align}
\end{itemize}

\begin{equation}
|M_u| =
\begin{pmatrix}
0.00136839 & 0.0770365 & 1.19782 \\
0.00778813 & 0.560874  & 2.93683 \\
0.24404    & 8.1122    & 147.741
\end{pmatrix} \,, \quad
\theta_u =
\begin{pmatrix}
 1.59621 &  2.80118  & -2.65001 \\
-2.34319 & -0.190895 & -0.644161 \\
-1.61289 &  0.584021 &  0.07447
\end{pmatrix}
\end{equation}

\begin{equation}
\rho_u =
\begin{pmatrix}
1.52494  & 1.57620  & 1.56165 \\
0.765990 & 1.01280  & 0.337923 \\
1.46664  & 0.895098 & 1.03875
\end{pmatrix} \,, \quad
\phi_u =
\begin{pmatrix}
1.39426  & 1.49660  & 1.50005 \\
0.716676 & 0.984072 & 0.332161 \\
1.39602  & 0.884794 & 1.03875
\end{pmatrix}
\end{equation}

\begin{equation}
m^u_1 = 0.369308~\mrm{MeV} \,, \qquad
m^u_2 = 0.409125~\mrm{GeV} \,, \qquad
m^u_3 = 147.999~\mrm{GeV} \,.
\end{equation}

\begin{align}
M_d &=
\begin{pmatrix}
0.00205044 & 0.0096169 & 0.0025584 \\
0.00702768 & 0.0985925 & 0.0173996 \\
0.242765   & 2.33774   & 0.76264
\end{pmatrix} \,, \quad
\theta_d =
\begin{pmatrix}
-0.184947   &  2.04673 &  1.12293 \\
-1.04910    &  1.68206 & -2.47164 \\
 0.00506372 & -2.31542 &  3.06043
\end{pmatrix}
\end{align}

\begin{equation}
\rho_d =
\begin{pmatrix}
1.07943  & 0.555546 & 0.583531 \\
0.326515 & 0.502659 & 0.350252 \\
0.689207 & 0.728280 & 0.938063
\end{pmatrix} \,, \quad
\phi_d =
\begin{pmatrix}
0.993588 & 0.522050 & 0.541247 \\
0.307557 & 0.483365 & 0.332447 \\
0.660453 & 0.712475 & 0.905822
\end{pmatrix}
\end{equation}

\begin{equation}
m^d_1 = 2.25527~\mrm{MeV} \,, \qquad
m^d_2 = 47.9153~\mrm{MeV} \,, \qquad
m^d_3 = 2.47254~\mrm{GeV} \,.
\end{equation}

\begin{itemize}
\item Configuration~II:
\begin{align}
c_Q &= \{0.628524, 0.546221, 0.285007\} \,, \notag \\
c_U &= \{-0.662224, -0.550397, 0.0801805\} \,, \notag \\
c_D &= \{-0.579521, -0.628656, -0.626738\} \,.
\end{align}
\end{itemize}

\begin{align}
M_u &=
\begin{pmatrix}
0.000705160 & 0.0296351 & 1.25154 \\
0.00391734  & 0.303462  & 4.75543 \\
0.157250    & 8.57855   & 148.068
\end{pmatrix} \,, \quad
\theta_u =
\begin{pmatrix}
3.03996 &  0.107148 &  2.03582 \\
1.79158 & -1.76781  & -2.88842 \\
2.07507 & -0.648895 &  3.02998
\end{pmatrix}
\end{align}

\begin{equation}
\rho_u =
\begin{pmatrix}
0.576867 & 0.721635 & 1.44261 \\
0.259565 & 0.598525 & 0.443977 \\
0.908033 & 1.47451  & 1.20473
\end{pmatrix} \,, \quad
\phi_u =
\begin{pmatrix}
3.03996 &  0.107148 &  2.03582 \\
1.79158 & -1.76781  & -2.88842 \\
2.07507 & -0.648895 &  3.02998
\end{pmatrix}
\end{equation}

\begin{equation}
m^u_1 = 1.05432~\mrm{MeV} \,, \qquad
m^u_2 = 0.399582~\mrm{GeV} \,, \qquad
m^u_3 = 148.398~\mrm{GeV} \,.
\end{equation}

\begin{align}
M_d &=
\begin{pmatrix}
0.00418127 & 0.000860589 & 0.00186071 \\
0.0663893  & 0.0619168   & 0.0228064  \\
2.43751    & 0.183510    & 0.140323
\end{pmatrix} \,, \quad
\theta_d =
\begin{pmatrix}
-2.87101 &  1.39416 & 2.70561 \\
-2.90716 &  1.17362 & 2.90809 \\
-1.80892 & -2.30267 & 2.33582
\end{pmatrix}
\end{align}

\begin{equation}
\rho_d =
\begin{pmatrix}
0.237597 & 0.231857 & 0.470886 \\
0.305562 & 1.35114  & 0.467477 \\
0.977694 & 0.348985 & 0.250662
\end{pmatrix} \,, \quad
\phi_d =
\begin{pmatrix}
-2.87101 &  1.39416 & 2.70561 \\
-2.90716 &  1.17362 & 2.90809 \\
-1.80892 & -2.30267 & 2.33582
\end{pmatrix}
\end{equation}

\begin{equation}
m^d_1 = 1.41124~\mrm{MeV} \,, \qquad
m^d_2 = 66.9487~\mrm{MeV} \,, \qquad
m^d_3 = 2.44931~\mrm{GeV} \,.
\end{equation}

\begin{itemize}
\item Configuration~III:
\begin{align}
c_Q &= \{0.627322, 0.570679, 0.272429\} \,, \notag \\
c_U &= \{-0.517935, -0.664365, 0.180466\} \,, \notag \\
c_D &= \{-0.576159, -0.610047, -0.638422\} \,,
\end{align}
\end{itemize}

\begin{align}
M_u &=
\begin{pmatrix}
0.147921 & 0.00223583 & 0.70694 \\
0.787783 & 0.00477027 & 4.06577 \\
8.66604  & 0.201339   & 145.112
\end{pmatrix} \,, \quad
\theta_u =
\begin{pmatrix}
-2.80680 &  2.86302   &  2.43167 \\
-0.23652 & -1.20710   & -1.23730 \\
 1.00216 &  0.0966827 &  0.0  
\end{pmatrix}
\end{align}

\begin{equation}
\rho_u =
\begin{pmatrix}
1.53467  & 1.88939  & 0.723485 \\
1.38068  & 0.680969 & 0.702896 \\
0.641530 & 1.21401  & 1.05965
\end{pmatrix} \,, \quad
\phi_u =
\begin{pmatrix}
-2.80680  &  2.86302   &  2.43167 \\
-0.236520 & -1.20710   & -1.23730 \\
 1.00216  &  0.0966827 &  0.0 
\end{pmatrix}
\end{equation}

\begin{equation}
m^u_1 = 1.49993~\mrm{MeV} \,, \qquad
m^u_2 = 0.553929~\mrm{GeV} \,, \qquad
m^u_3 = 145.430~\mrm{GeV} \,.
\end{equation}

\begin{align}
M_d &=
\begin{pmatrix}
0.0122178 & 0.00379117 & 0.00346894 \\
0.0813964 & 0.0316802  & 0.0033306  \\
2.33248   & 0.899976   & 0.488706
\end{pmatrix} \,, \quad
\theta_d =
\begin{pmatrix}
2.54815  &  2.37217  & -1.79028 \\
0.769324 & -0.385483 &  0.262617 \\
0.348142 &  2.10335  &  0.0
\end{pmatrix}
\end{align}

\begin{equation}
\rho_d =
\begin{pmatrix}
0.603011 & 0.537917 & 1.23789 \\
0.678640 & 0.759331 & 0.200775 \\
0.821415 & 0.911140 & 1.24436
\end{pmatrix} \,, \quad
\phi_d =
\begin{pmatrix}
2.54815  &  2.37217  & -1.79028 \\
0.769324 & -0.385483 &  0.262617 \\
0.348142 &  2.10335  &  0.0 
\end{pmatrix}
\end{equation}

\begin{equation}
m^d_1 = 2.38820~\mrm{MeV} \,, \qquad
m^d_2 = 60.8655~\mrm{MeV} \,, \qquad
m^d_3 = 2.54821~\mrm{GeV} \,.
\end{equation}

\section{\label{app:EWdF} Electroweak contributions to the tree-level
$\Delta F = 2$ FCNCs}
The electroweak contributions to $\Delta F = 2$ FCNCs come from the tree-level
processes mediated by the KK photons, the $Z$ boson, and the heavy $Z'$ boson
that arise due to the $SU(2)_R$ in the MCRS model~\cite{CusRS}~\footnote{Note
unlike the $Z$ field which has [+,+] boundary conditions, the $Z'$ field has
[-,+] boundary conditions, which give rise to KK excitations only.}. As we show
below, the electroweak contributions are small due to the suppression of
the (much) smaller electroweak interaction strength relative to that of the
strong interaction (at the NP scale $\Lambda$).

The electroweak gauge bosons contribute to the $\Delta F = 2$ processes in two
ways. They contribute either directly through the four-fermion process
analogous to that in Fig.~\ref{Fig:gKK4f}, or they modify the gauge-fermion
vertex through mixings with gauge and fermion KK modes as discussed in
Ref.~\cite{CNW08}. In the former case, all electroweak gauge KK modes can
contribute, while the latter only happens via the mixing of the $Z$ zero mode
with the $Z'$ KK modes and the KK fermions.

For the direct electroweak contribution, the Wilson coefficients have similar
forms as those for the KK gluons given in Eq.~\eqref{Eq:WCs}, but with
appropriate changes in the numerical coefficients (no colour factor $1/3$),
the interaction strengths, and gauge boson wavefunctions:
\begin{align}
C_1(\Lambda)^{EW} &= \frac{1}{2}\mathfrak{S}^{LL}_{ab,ab}(A_{KK}) \,, &
\wtil{C}_1(\Lambda)^{EW} &= \frac{1}{2}\mathfrak{S}^{RR}_{ab,ab}(A_{KK}) \,,
\notag \\
C_4(\Lambda)^{EW} &=  0 \,, &
C_5(\Lambda)^{EW} &= -2\mathfrak{S}^{LR}_{ab,ab}(A_{KK}) \,,
\end{align}
where $A = \gamma,\,Z,\,Z'$. Note that without the colour structure, there is
no electroweak contribution to the Wilson coefficient $C_4$ at tree-level.

For the photon and the $Z$ (and their respective KK excitations), their
couplings to the fermions are the same as in the SM. For the $Z'$, its
coupling to the fermions depends on $g_R$, the gauge coupling constant of
$SU(2)_R$. Since it is commonly assumed in the literature that the coupling
constants of $SU(2)_{L,R}$ are equal, for the purpose of comparison we take
$g_R = g_L$ also. The $Z'ff$ coupling is then given by $g_{Z'}Q_{Z'}(f)$,
where
\begin{equation}
g_{Z'} = \frac{c\,g_Z}{\sqrt{1-s^2/c^2}} \,, \qquad
Q_{Z'}(f) = T^3_R(f) - \frac{s^2}{c^2}\frac{Y_f}{2} \,,
\end{equation}
with the usual definitions $g_Z = e/(s c)$, $s = e/g_L$, and
$c = \sqrt{1-s^2}$.

In Table~\ref{Tb:WCEW}, we list the ratio of electroweak contribution to the
Wilson coefficients at $\Lambda = 4$~TeV to that due to KK gluons alone for
each source of the direct electroweak tree-level four-fermion processes.
\begin{table}[htbp]
\begin{ruledtabular}
\begin{tabular}{cccccc}
Wilson Coefficient Ratio & Fermion Type & KK $\gamma$ & KK $Z$ & KK $Z'$ & Total \\
\hline
$C_1(\Lambda)^{EW}/C_1(\Lambda)^{QCD}$ & u & 0.13  & 0.18 & 0.0054 & 0.32 \\
            & d & 0.033 & 0.28 & 0.0054 & 0.32 \\
$\wtil{C}_1(\Lambda)^{EW}/\wtil{C}_1(\Lambda)^{QCD}$ & u & 0.13  & 0.044 & 0.14 & 0.32 \\
               & d & 0.033 & 0.011 & 0.28 & 0.32 \\
$C_5(\Lambda)^{EW}/C_5(\Lambda)^{QCD}$ & u & -0.27  & 0.18 &  0.056 & -0.033 \\
            & d & -0.067 & 0.11 & -0.077 & -0.033
\end{tabular}
\end{ruledtabular}
\caption{\label{Tb:WCEW} Ratio of direct electroweak contributions to KK gluon
contributions in the Wilson coefficients. The fermion type ``u'' (``d'') denotes
that up-type (down-type) quarks are involved in the $\Delta F = 2$ FCNC process.
All quarks have SM quantum numbers.}
\end{table}
Note that because the difference between the overlap integrals for the [+,+]
and [-,+] gauge bosons is very small as the respective bulk profiles are almost
the same in regions where large overlap happens, the ratios listed in 
Table~\ref{Tb:WCEW} are essentially just that of the respective charge factors 
and electroweak gauge coupling constants. Note also that the total electroweak
contribution for up-type and down-type quarks are the same, and that for $C_1$
and $\wtil{C}_1$ are the same. This can be seen most easily in the gauge 
interaction basis where the magnitude of the gauge charges are the same for 
both up-type and down-type quarks, and the $SU(2)_{L,R}$ quark quantum numbers
are the same. Since the direct processes depend on the square of the gauge 
charges, the conclusion follows.

We remark here that the contributions to the Wilson coefficients due to KK 
gluon and KK photon are universal for all RS models with bulk fermions, but 
those due to $Z$ and $Z'$ are not. This is because the coupling of the $Z$ 
and $Z'$ to fermions depend on the representation in which the fermions are
embeddd in the gauge group of the model. Throughout this work and in 
Table~\ref{Tb:WCEW}, the bulk fermions are embedded such that the SM LH 
doublets (singlets) are $SU(2)_R$ singlets (doublets) so that they have SM 
quantum numbers (see Eq.~\eqref{Eq:qrep}). However, ratios different from 
those listed in Table~\ref{Tb:WCEW} would arise if different fermion 
representation is used. For example, we list the electroweak to KK gluon 
ratios in Table~\ref{Tb:WCEW2} in the case where the SM LH doublets are 
embedded as bifundamentals in the $SU(2)_L \times SU(2)_R$, and the up-type 
(down-type) singlets as $SU(2)_R$ singlets (triplets) (see e.g 
Ref.~\cite{DF07}) so that there is a left-right parity~\cite{ACRP06}.
\begin{table}[htbp]
\begin{ruledtabular}
\begin{tabular}{cccccc}
Wilson Coefficient Ratio & Fermion Type & KK $\gamma$ & KK $Z$ & KK $Z'$ & Total \\
\hline
$C_1(\Lambda)^{EW}/C_1(\Lambda)^{QCD}$ & u & 0.13  & 0.18 & 0.36 & 0.67 \\
            & d & 0.033 & 0.28 & 0.56 & 0.87 \\
$\wtil{C}_1(\Lambda)^{EW}/\wtil{C}_1(\Lambda)^{QCD}$ & u & 0.13  & 0.044 & 0.087 & 0.26 \\
               & d & 0.033 & 0.011 & 1.43  & 1.47 \\
$C_5(\Lambda)^{EW}/C_5(\Lambda)^{QCD}$ & u & -0.27  & 0.18 &  0.35 &  0.26 \\
            & d & -0.067 & 0.11 & -1.78 & -1.74
\end{tabular}
\end{ruledtabular}
\caption{\label{Tb:WCEW2} Ratio of direct electroweak contributions to KK gluon
contributions in the Wilson coefficients. The fermion type ``u'' (``d'') denotes
that up-type (down-type) quarks are involved in the $\Delta F = 2$ FCNC process.
Here the SM LH doublets, and u-type and d-type singlets transform as
$(2,\bar{2})_{2/3}$, $(1,1)_{2/3}$, and $(1,3)_{2/3}$ under
$SU(2)_L \times SU(2)_R \times U(1)_X$ respectively.}
\end{table}
Note that only the KK $Z'$ contributions are different in changing to this
representation because only $Q_Z'$ is sensitive to the different assignment of
the $SU(2)_R$ quantum numbers.

For the electroweak contributions due to mixings, the effects are no longer
universal -- the suppression factors are no longer determined by the
electroweak charges and coupling constants alone -- as there is now dependence
on the fermion localization parameters and the quark mixing matrices. However,
they are generically expected to be small as they are
$\mathcal{O}(v^4/\Lambda^4)$ compared to the direct contributions~\footnote{The
zero mode-KK mode mixing happens through interactions with the Higgs, hence
the effective coupling of the fermions to the zero mode of $Z$ is
$\mathcal{O}(v^2/\Lambda^2)$ compared to the direct fermion couplings to the KK
modes of $Z$ and $Z'$. More details can be found in Ref~\cite{CNW08}.},
although non-generic enhancement may happen depending on the particular quark
mixing matrices involved, which typically do not exceed $\mathcal{O}(0.01)$ of
the KK gluon contributions. We have checked in each case that the combined
effect of the direct and mixing electroweak contributions do not appreciably
alter the KK gluon contributions to the Wilson coefficients, and that they are
still well within the UTfit bounds.

\end{document}